\documentclass[conference]{IEEEtran}
\IEEEoverridecommandlockouts
\usepackage{cite}
\usepackage{amsmath,amssymb,amsfonts}
\usepackage{algorithmic}
\usepackage{graphicx}
\usepackage{siunitx}
\usepackage{textcomp}
\usepackage{xcolor}
\usepackage{svg}
\usepackage{balance}
\usepackage{placeins} 
\def\BibTeX{{\rm B\kern-.05em{\sc i\kern-.025em b}\kern-.08em
    T\kern-.1667em\lower.7ex\hbox{E}\kern-.125emX}}

\begin{document}

\title{High-Density MIMO Localization Using a 32×64 Ultrasonic Transducer-Microphone
Array with Real-Time Data Streaming}

\author{\IEEEauthorblockN{
Rens Baeyens\IEEEauthorrefmark{1}\IEEEauthorrefmark{2},
Dennis Laurijssen\IEEEauthorrefmark{1}\IEEEauthorrefmark{2},
Jan Steckel\IEEEauthorrefmark{1}\IEEEauthorrefmark{2} and
Walter Daems\IEEEauthorrefmark{1}\IEEEauthorrefmark{2}}
\IEEEauthorblockA{\IEEEauthorrefmark{1}FTI Cosys-Lab,
University of Antwerp,
Antwerp, Belgium, Email: walter.daems@uantwerpen.be}
\IEEEauthorblockA{\IEEEauthorrefmark{2}Flanders Make Strategic Research Centre
Lommel, Belgium}}

\maketitle

\begin{abstract}
In this work, we present a novel ultrasonic array system designed for high-precision localization using a large-scale MIMO (Multiple-Input Multiple-Output) architecture. The system combines 32 transmitters with 62 microphones, creating an extended virtual aperture that improves channel separability and spatial resolution. Each transmitter is excited by a random-phase multisine within the ultrasonic band, which reduces inter-channel correlation and increases robustness against multipath. The feasibility of the approach is demonstrated through simulations of reflector imaging and analysis of channel separation under realistic transducer bandwidth constraints. Results show that MIMO processing enables improved separation of reflectors compared to single-emitter configurations, although practical limitations such as transducer bandwidth reduce the achievable channel isolation.
\end{abstract}


\section{Introduction}
Ultrasonic sensing remains a critical modality for short-range localization in environments where optical and RF-based methods face limitations due to harsh conditions such as occlusion, multipath, or power constraints. However, conventional ultrasonic systems that are typically based on mono-static or narrow-aperture array configurations, are fundamentally constrained in spatial resolution, angular coverage, and robustness to environmental complexity~\cite{Kerstens2019,Jansen2024,Izquierdo2024,Allevato2021,Haugwitz2022}. These limitations are particularly pronounced in applications requiring fine-grained localization such as indoor navigation, robotic mapping, or gesture tracking~\cite{Verellen2020,Kerstens2023,Schenck2025,Athi2009,Gomes2024}. In this work, we present a broadband ultrasonic localization platform based on a large-scale Multiple-Input Multiple-Output (MIMO) array, comprising 32 transmitters and 64 microphones. This architecture significantly expands the system’s virtual aperture and spatial sampling density, enabling rich directional diversity in the received wavefield~\cite{Kerstens2019,STSTAB,}. To overcome the bandwidth limitations of traditional 40kHz transducers, we employ Knowles SPH0641 PDM digital microphones, which support broadband ultrasonic sensing with improved linearity and signal-to-noise ratio (SNR). This enables extraction of finer temporal and spectral features, enhancing the localization accuracy and robustness. The hardware platform contains an STM32F429 microcontroller to generate ultrasonic transmissions and a Raspberry Pi RP2350B microcontroller to manage concurrent multi-channel acquisition via its Programmable I/O (PIO) subsystem, supporting scalable digital microphone interfacing without overloading CPU resources. Data is streamed in real time to a host system via a second PIO-driven interface connected to an FT232H USB FIFO bridge, potentially enabling sustained acquisition and transmission for up to 64 channels. In the current stage of development, we have successfully achieved continuous acquisition and data transfer for up to 16 microphones. The bottleneck that limits the number of simultaneously captured and transmitted channels is discussed in Section II-A2. To validate the system’s potential, we present simulation results in representative localization scenarios. Specifically, we analyze channel separation under realistic transducer bandwidth constraints and demonstrate through simulated reflector imaging that MIMO processing provides superior separation compared to single-emitter configurations~\cite{Kerstens2019,STSTAB,Kerstens2023}. These results highlight both the advantages of the large-scale MIMO approach and the practical limitations imposed by hardware bandwidth. Future work will extend the platform with real-time beamforming, including MVDR methods, to further improve spatial accuracy and robustness under multipath conditions~\cite{Job2023,Wirdhani2025,Islam2024}. This work establishes a scalable, low-cost platform for high-resolution ultrasonic sensing, with potential applications in robotics, human–computer interaction, and smart environments.

\begin{figure*}[h!]
\centering
    \includegraphics[width=0.9\textwidth]{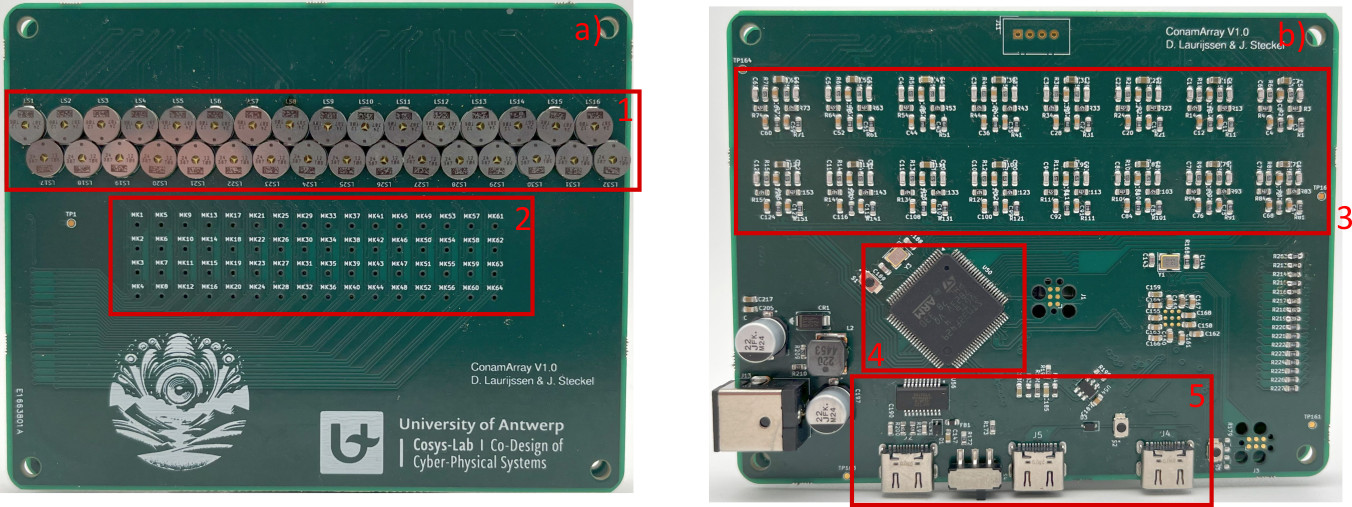}
    \caption{The developed ConamArray hardware platform with the front-end (panel a) containing 32 Conamara UA-C0603-3T transducers (box 1) and 64 SPH0641LU4H-1 MEMS microphones (box 2) arranged in a 4 by 16 uniform rectangular array. The back-end PCB (panel b) contains the transducer driving circuitry, the two microcontrollers (panels 4 and 5), and an FT232H usb bridge (panel 5). The ConamArray has a footprint of $102\times{}80\unit{\milli\meter}$.}
    \label{fig:syshardware}
\end{figure*}

\section{Methods}
\subsection{System Architecture}
\label{system_arch}
\subsubsection{Embedded Hardware}
\label{hardware}
\begin{figure}[ht]
\centering
    \includegraphics[width=0.9\linewidth]{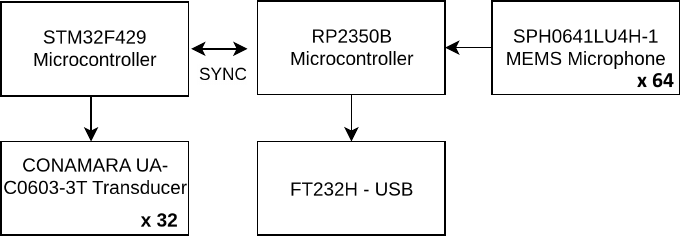}
    \caption{High-level overview of the ConamArray system architecture, showing the two subsystems: one for the emitter, powered by an STM32 microcontroller, and one for the receiver array, powered by an RP2350B microcontroller. Data transfer from the RP2350B habbens through a FT323H USB FIFO interface. Both subsystems are synchronized by a LVTTL synchronization interface.}
    \label{fig:sysarch}
\end{figure}

The high-level overview of the hardware architecture is presented in Figure~\ref{fig:sysarch}. Acquisition and speaker control are divided into two separate microcontrollers mainly due to the large amount of GPIOs necessary to achieve control and acquisition of the substantial amount of transducers and microphones. The driver circuits for the 32 USound Conamara UA-C0603-3T transducers are controlled by the STM32F429 microcontroller, whereas the RP2350B microcontroller is responsible for synchronous acquisition of the 64 MEMS microphones. This acquired data is then forwarded to the FT232H USB Bridge that has a theoretical maximal datarate of 40 MBytes/sec. The systems share a GPIO line that is used as synchronization/trigger signal. 

\subsubsection{Embedded Firmware}
\label{firmware}
The STM32F429 drives the transducers in a synchronous manner for the 32 channels. Each transmit channel is excited by a length-N zero-mean random-phase signal, band-limited by a digital filter to the [20–80] kHz band. The resulting waveform approximates a random-phase multisine by concentrating power in the desired band. These signals are chosen to increase the received signal's variability and partially prevent particular echoing blind spots. The signal model of the transmit sequences is described in detail in Section \ref{multisinetransmit}.
The RP2350B microcontroller is employed to manage real-time acquisition of acoustic data from an array of microphones using a Programmable I/O (PIO) module. The microphones generate pulse density modulated (PDM) signals, which are synchronously sampled at $4.5\unit{\mega\hertz}$. The resulting data is aggregated into acoustic frames with a configurable frame size, and transferred via Direct Memory Access (DMA) to a second PIO module. This module synchronizes with the clock of an FT232H USB bridge to facilitate low-latency, continuous streaming of data to a host system over USB.

While the FT232H supports a theoretical maximum throughput of $40\unit{MB\per\second}$, the current implementation achieves only 50\% of this bandwidth due to the usage of only one of the two available FIFO transmission slots, effectively limiting the transfer rate to $20\unit{MB\per\second}$.
Further constraints arise from a blocking mechanism on the host PC. During periods of high system load or USB bus contention, the FT232H asserts flow control by pulling the TX enable line low, signaling the RP2350B to pause transmission. When this occurs, the host system temporarily ceases to process incoming data, posing a risk of data loss. To mitigate this, the TX enable line is continuously monitored by the RP2350B's PIO logic, ensuring that data transmission is halted when necessary and resumed when the host becomes available. To maintain system stability under these constraints, the number of active microphones is currently capped at 16. However, preliminary testing indicates that up to 32 microphones can be supported if the host-side blocking is more effectively controlled. The RP2350B has already been overclocked to $250\unit{\mega\hertz}$ to match the timing requirements imposed by synchronization with the FT232H’s clock, highlighting the tight performance margins.
Achieving the system’s full potential of 64 simultaneously sampled microphones would require full utilization of both FIFO transmission slots, thereby restoring access to the full $40\unit{MB\per\second}$ bandwidth. This would also necessitate improvements in host-side USB handling and further optimization of the synchronization mechanism used to synchronize with the FT232H clock.

\begin{figure*}[t!]
\centering
    \includegraphics[width=0.9\textwidth]{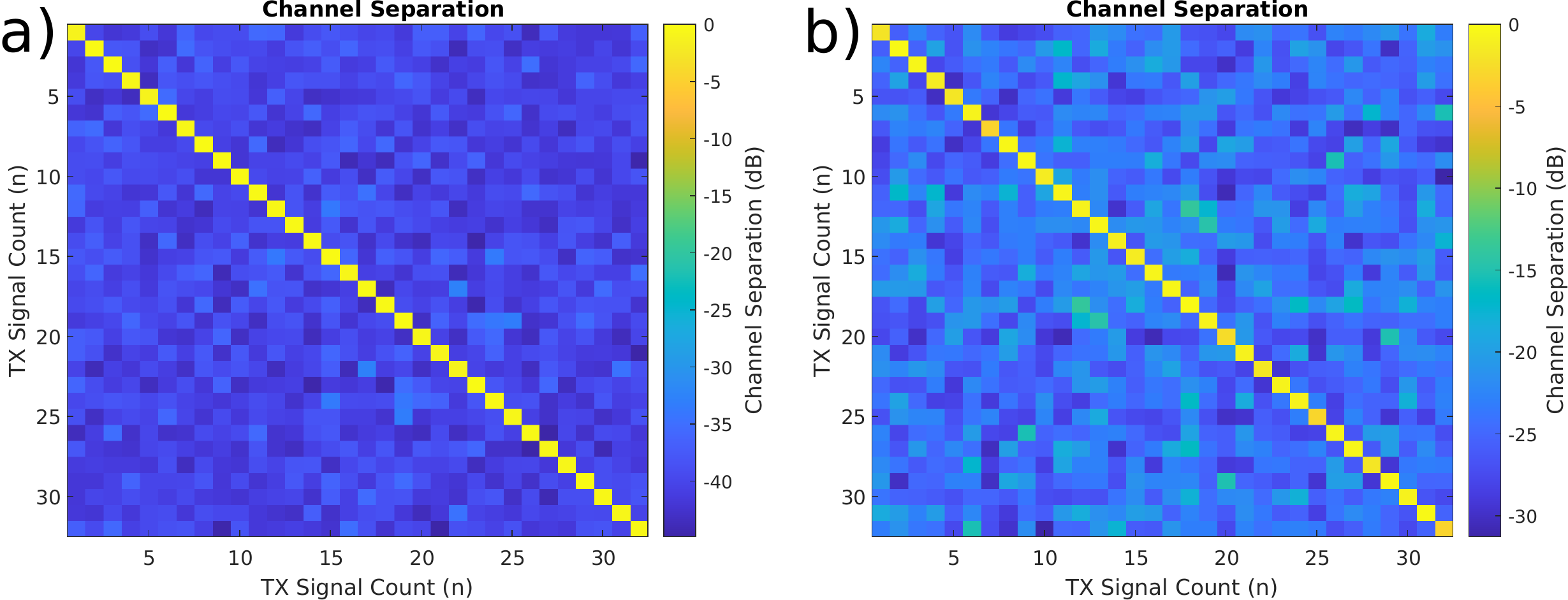}
    \caption{Channel separation in dB between the 32 generated random-phase multisine sequences is calculated using cross correlation. In this figure, we plot the maximum of the cross correlation function, between all the differerent emission signals. Panel a) shows tge result of this computation for a theoretical bandwidth between 20kHz and 80kHz. Panel b) shows the result for this calculation with the measured speaker response, which is not flat as can be seen in Figure \ref{fig:transferfunction}}.
    \label{fig:channel_separation}
\end{figure*}

\subsection{Signal Processing}
\subsubsection{Transmitted Signals}
\label{multisinetransmit}
Each of the 32 transducers in the ConamArray system is driven by an independent, band-limited random-phase multisine signal~\cite{Vanhoenacker2003,Hoagg2006}. For transmit channel $c = 1, \dots, 32$, the signal can be expressed as

\begin{equation}
x_c(t) = \sum_{k=1}^{K} A_k \cos(2 \pi f_k t + \phi_{c,k}),
\end{equation}

where $f_k$ are the discrete frequency components within the [38--42~kHz] band, $A_k$ are the amplitudes, and $\phi_{c,k}$ are independent random phases for each channel. The random phases reduce inter-channel correlation and mitigate coherent reflections, increasing the effective spatial diversity of the transmitted wavefield. All 32 sequences are transmitted simultaneously, enabling full MIMO operation and maximizing the system's virtual aperture for high-resolution localization.

\subsubsection{Received Signals}
The transmitted sequences are then recorded by an array of $K=64$ microphones. The recorded signals $r_k[n]$ at each microphone $k$ can be modeled as:
\begin{equation}
\mathbf{r}_k[n] = \sum_{i=1}^{M} \big( \mathbf{h}_{ik}[n] \ast \mathbf{x}_i[n] \big) + \mathbf{v}_k[n],
\end{equation}

where $\mathbf{h}_{ik}[n]$ is the impulse response from transmitter $i$ to microphone $k$,  and $\mathbf{v}_k[n]$ represents additive noise.

\begin{figure}[h!]
\centering
    \includegraphics[width=0.9\linewidth]{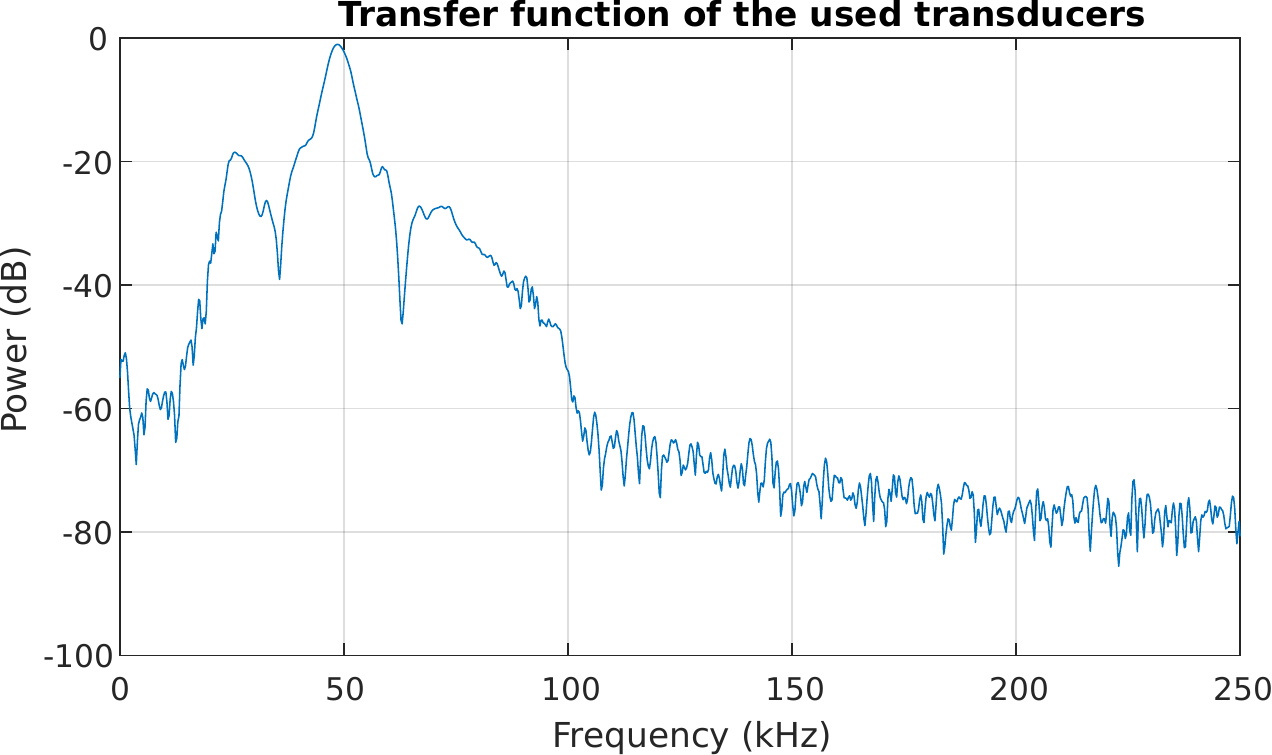}
    \caption{Transfer function of the used transducer when being stimulated with a multisine sequence, bandlimited around 20kHz to 80kHz. Significant dips can be seen in the emitted spectrum, which is in accordance with the datasheet of the transducer.}
    \label{fig:transferfunction}
\end{figure}

\begin{figure*}[t!]
\centering
    \includegraphics[width=0.9\textwidth]{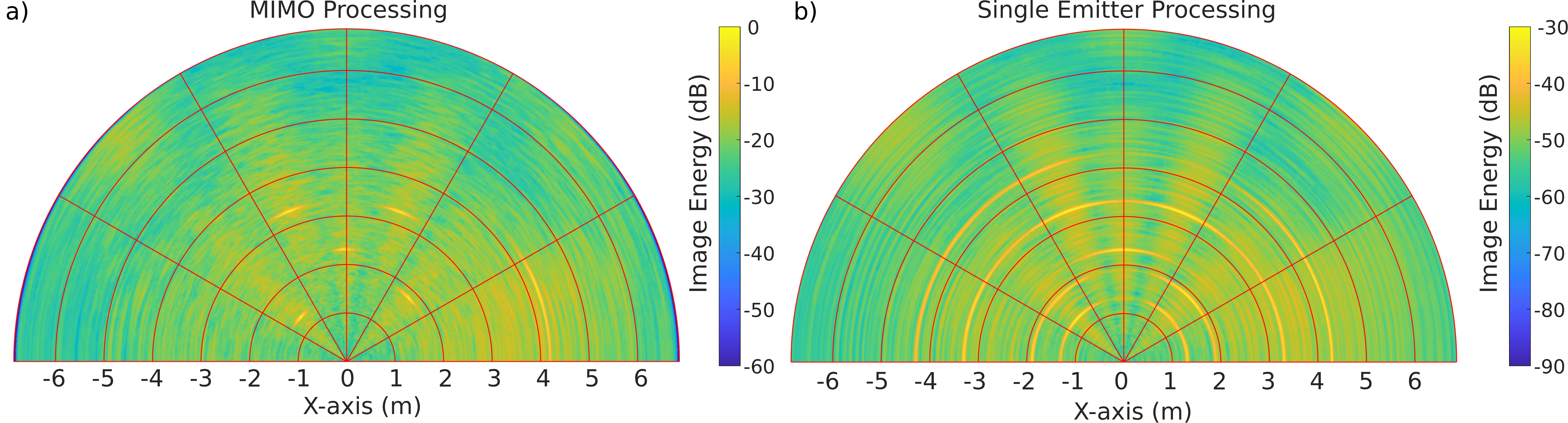}
    \caption{Simulation of imaging for the separation of 6 reflectors using single emitter processing (panel b) and MIMO processing (panel a). A significant increase in spatial accuracy can be observed when moving from a single emitter system to a MIMO processing approach, similar to the results in~\cite{STSTAB}. The same increase can be observed for the peak-to-sidelobe level. Finally, the total image strength is increased by 30dB when moving from a single to the 32 emitter setup.}
    \label{fig:imaging_separation}
\end{figure*}

Each transmitted sequence $\mathbf{x}_i[n]$ was extracted independently from the composite microphone signals using matched filtering~\cite{Weinstein1993}:

\begin{equation}
\mathbf{y}_{ik}[n] = \mathbf{r}_k[n] \star \mathbf{x}_i[-n],
\end{equation}

The matched filter maximizes the signal-to-noise ratio for the corresponding sequence while minimizing interference from other sequences. Outputs from all microphones were then aligned and normalized by the sequence energy, enabling reconstruction of the transmitted sequences and characterization of the acoustic channels.

\section{Preliminary Results}
To validate the theoretical feasibility, the channel separation between the generated sequences from~\ref{multisinetransmit} is calculated and presented in Figure~\ref{fig:channel_separation}. Panel a) presents the ideal scenario without bandwidth limitations, whereas b) takes the Conamara speaker bandwidth into account. A significant difference in channel separation is introduced by the bandwidth limitation.

The benefits in imaging capability when using MIMO compared to a single emitter setup are simulated in Figure~\ref{fig:imaging_separation}. The simulation is elaborated for a setup with 6 reflectors in range. The left panel of Figure~\ref{fig:imaging_separation} presents the improved imaging separation of the reflectors compared to the right panel, using a single emitter for image separation.

\section{Discussion}
The preliminary results demonstrate the potential of large-scale MIMO ultrasonic systems for achieving fine-grained localization accuracy. The proposed 32$\times$64 array significantly increases the spatial diversity and virtual aperture size compared to conventional narrow-aperture or mono-static arrays. The random-phase multisine transmit strategy ensures low inter-channel correlation, which is crucial for reliable channel separation and robust matched filtering. Simulations confirm that MIMO operation yields substantially improved imaging and separation performance, even in multipath-rich environments.

Despite these promising results, several system-level limitations remain. The current acquisition pipeline is limited to 16 microphones due to USB bandwidth constraints and host-side flow control blocking. This bottleneck prevents full exploitation of the available 64 microphones, which would otherwise provide higher spatial resolution and robustness. Furthermore, real-time beamforming was not yet implemented on the embedded platform. Instead, simulations were used to estimate the system’s capabilities. This highlights the need for efficient on-board processing to achieve true real-time localization. Finally, while broadband MEMS microphones extend the usable frequency range, transducer bandwidth still imposes constraints on effective channel separation and achievable spatial resolution.

\section{Future Work}
Future work will focus on both hardware and algorithmic improvements to unlock the full potential of the ConamArray platform. On the hardware side, achieving simultaneous acquisition from all 64 microphones will require full utilization of the FT232H dual-FIFO interface and optimization of host-side USB drivers to sustain the theoretical $40\unit{MB/s}$ throughput. Alternative high-speed data interfaces, such as USB 3.0 or Gigabit Ethernet, may also be considered to alleviate current streaming bottlenecks. In parallel, dedicated FPGA-based acquisition could provide additional scalability compared to the current microcontroller-driven approach.
On the signal processing side, the real-world evaluation of the signal separability and the integration of real-time beamforming algorithms on embedded hardware will be primary targets. Low-complexity approximations of MVDR or subspace-based methods will be investigated to balance computational load with localization accuracy. Extending the system to dynamic scenarios, such as moving targets or robotic navigation, will provide further validation of the robustness of the MIMO architecture.

\bibliographystyle{IEEEtran}
\bibliography{fulllib}

\end{document}